# The gravitational Field of a massless Particle on the Horizon of a stationary Black Hole

Albert Huber[*]

UAS Technikum Wien - Department Applied Mathematics and Physics, Höchstädtplatz 6, 1200 Vienna, Austria

**Abstract**

In this work, the field of a gravitational shockwave generated by a massless point-like particle is calculated at the event horizon of a stationary Kerr-Newman black hole. Using the geometric framework of generalized Kerr-Schild deformations in combination with the spin-coefficient formalism of Newman and Penrose, it is shown that the field equations of the theory, at the event horizon of the black hole, can be reduced to a single linear ordinary differential equation for the so-called profile function of the geometry. This differential relation is solved exactly. Based on the results obtained, a physical interpretation is given for the found shockwave spacetime, and it is clarified how these results lead back to those of previous works on the subject, which deal with the much simpler cases of gravitational shockwaves in static black hole backgrounds.

## Introduction

The problem of calculating the gravitational field of a massless ultrarelativistic particle which propagates on a flat Minkowskian background was solved a long time ago by Aichelburg and Sexl [1]. Leading to the metric of a deformed singular spacetime geometry of an ultrarelativistic point mass, the solution of this problem was found by performing a Lorentz boost of the Schwarzschild geometry in isotropic coordinates and executing an appropriate ultrarelativistic limit. Ever since the discovery of the corresponding geometry, numerous approaches have focused on providing similar solutions to Einstein's equations for more general geometric settings and broader scopes of physical application. These approaches include, in particular, the works of Dray and 't Hooft [17] and Sfetsos [32], which provide the precise form of a gravitational shock wave caused by a massless particle in static black hole and cosmological backgrounds.

---

[*]hubera@technikum-wien.at

Describing physically the gravitational field of an ultrarelativistic point-like source, which represents the field of a particle whose rest mass becomes zero in the limit previously mentioned, the said shock wave geometries share a prodigious mathematical property: They are characterized by metrics that are distributionally valued and therefore contain contributions of uncharacteristically low regularity from the point of view of classical general relativity. These low regularity contributions occur as a byproduct of a specific distributional method known as Penrose's 'scissors-and-paste' procedure [29], which has been applied to some well-known spherically symmetric gravitational backgrounds, including the spacetimes of Schwarzschild, Reissner-Nordström and de Sitter.

The underlying method, which is based on the idea of performing a specific coordinate shift in one of the components of the metric in double null coordinates, entails the existence of a confined particle-like source that generates a gravitational shock wave that skims along the event horizon of the black hole. This involves the introduction of a delta-shaped profile alias shift function (which is actually a distribution) that has a compact support on a single null hypersurface of spacetime. The reduced version of this profile or shift function represents the only unknown of the theory and therefore ought to be determined by explicitly solving the field equations of the theory.

However, this usually involves serious mathematical problems arising from the fact that the modifications caused by the field of a point-like particle require performing nonlinear operations on distributions, which are generally ill-defined. This in turn manifests itself in the fact that the nonlinear curvature fields associated with the deformed spacetime metric typically contain highly problematic, distributionally ill-defined 'squares' of the delta distribution, which cannot be treated meaningfully from a mathematical point of view and therefore were recklessly ignored in practice.

A notable exception to this approach is a work by Alonso and Zamorano [2], which showed (for the special case of Reissner-Nordström spacetime) that the corresponding lack of mathematical rigour can elegantly be overcome by using a geometrically more appealing approach, known as the generalized Kerr-Schild framework. This framework, which makes it possible to rigorously deal with quantities of low regularity, is based on performing a specific null geometric deformation of spacetime, commonly referred to as a generalized Kerr-Schild ansatz, which forms the basis of the so-called generalized Kerr-Schild framework.

Due to its linearity, this geometric framework is tailor-made for dealing with the low regularity of the components of the deformed field equations. More precisely, because of the fact that the mixed deformed Einstein tensor of the geometry (and hence also the mixed deformed Ricci) is always linear in the profile function, standard distribution theory can be used for solving the mixed Einstein equations with respect to a given background.

Based on this observation, a general approach to the problem was presented in [10], which uses the generalized Kerr-Schild framework, but does not rely on the consideration of a specific background geometry of spacetime. Despite this generality of the corresponding geometric setting, a closer examination of the generalized Kerr-Schild class of the Schwarzschild and Reissner-Nordström black hole metrics has shown that the said approach allows one to reproduce both the results of Dray and 't Hooft and Sfetsos locally by specifying the mixed deformed Einstein tensor of the respective static black hole backgrounds with respect to various components of the corresponding Riemann and Weyl curvature tensors. In addition, it was shown that the Aichelburg-Sexl geometry can be calculated as another special case of this setting.

In the present work, these results are to be extended to the much more complicated case of a gravitational shock wave produced by a massless particle-like source at the event horizon of a stationary, axially symmetric Kerr-Newman black hole. For this purpose, the Newman-Penrose spin-coefficient formalism is applied to the generalized Kerr-Schild framework, which allows one to expand the gravitational field equations in terms of spin-coefficients. As a result, an ordinary differential equation for the profile function of the geometry is derived, which is given with respect to a specific differential operator that coincides with the genericity operator for weakly isolated horizons of the theory of Ashtekar et al. and happens to be closely related to the MOTS stability operator of Andersson, Mars and Simon [3, 7]. The differential equation for the massless particle shock wave on the black hole event horizon defined by this differential operator is solved exactly in this work using methods of the theory of linear differential and difference equations with variable coefficients. This is achieved on the basis of the fact that the said relation belongs to the so-called Fuchsian class of second order differential equations with coefficients with five regular singular points, so that one immediately knows the exact form of the two linearly independent solutions, which can be superimposed to a single (distributionally defined) solution for the profile function of the geometry. By using Mallik's companion matrix approach [28], the explicit form of this solution is obtained in terms of infinite power series. In this way, a local geometric backreaction is specified, which is given by a superimposed Kerr-Schild ansatz. This backreaction generalizes the asymptotically flat stationary field of a Kerr-Newman black hole and causes a change in the energy-momentum tensor in the form of an ultrarelativistic particle field, thus providing a novel generalized Kerr-Schild geometry in four dimensions.

## 1 Einstein's Field Equations I: Deriving the generalized Dray-'t Hooft Relation

To determine the exact structure of the geometric field of an ultrarelativistic point-like particle in a stationary black hole background, the first step shall be the application of the mathematical approach worked out in [10] to the case of a Kerr-Newman black hole background. This background can be classified as a Petrov-type $D$ spacetime, which brings about a decisive simplification in that, according to the Goldberg-Sachs theorem, it means that the coefficients $\Psi_0$, $\Psi_1$, $\Psi_3$, $\Psi_4$ and $\phi_0$, $\phi_2$ of the Weyl and Einstein-Maxwell tensors can all be set to zero by the use of a suitable null geodesic frame [18, 27].

Consequently, in order to be able to deal with a maximally simplified geometric setting, it seems reasonable to consider the Kerr-Newman metric $g_{ab}$ in Kerr coordinates. As is well-known, the individual components of the Kerr-Newman metric in these coordinates can be read off from the line element

$$ds^2 = -(1 - \frac{2Mr - e^2}{\Sigma})dv^2 + 2(dv - a\sin^2\theta d\phi)dr + \Sigma d\theta^2 + \qquad (1)$$
$$+ \frac{\Pi \sin^2\theta}{\Sigma}d\phi^2 - \frac{2(2Mr - e^2)}{\Sigma}a\sin^2\theta dv d\phi,$$

where $\Sigma = r^2 + a^2\cos^2\theta$, $\Pi = (r^2 + a^2)^2 - \Delta a^2 \sin^2\theta$ and $\Delta = r^2 + a^2 - 2Mr + e^2$.

There are two main reasons for the given choice of Kerr coordinates: On the one hand, these coordinates are regular at the internal and external event horizons $r_{\pm} = M \pm \sqrt{M^2 - a^2 - e^2}$. On the other hand, they allow to directly read off the only two principal null directions of the geometry, which lead to associated null congruences that are globally shear free.

Sure enough, given this comparatively simple geometric setting, the decomposition relations

$$\tilde{g}_{ab} = g_{ab} + f l_a l_b, \qquad (2)$$

and

$$\tilde{g}^{ab} = g^{ab} - f l^a l^b \qquad (3)$$

can be formulated with respect to the background spacetime $(M, g)$ in order to obtain a new spacetime $(\tilde{M}, \tilde{g})$ with different metric and curvature components. In this context, $g_{ab}$ and $g^{ab}$ represent the metric and the inverse metric of the Kerr-Newman background, the scalar function $f$ represents the so-called profile function and $l_a$ represents a co-vector field, whose index can be raised and lowered with the background metric, so that in particular $l_a = g_{ab}l^b$ applies. In this context, the corresponding vector field is assumed to be null geodetic, which means that it must satisfy the relations $g_{ab}l^a l^b = 0$ and $Dl^a = 0$, where $D := l^b \nabla_b$ applies by definition. In that respect, the deformation relation (2) provides a new deformed metric belonging to the generalized Kerr-Schild class of the given Kerr-Newman black hole background.

With relations (2) and (3) at hand, the mixed Einstein tensor of generalized Kerr-Schild class

$$\tilde{G}^a_b = \tilde{R}^a_b - \frac{1}{2}\delta^a_b \tilde{R} \qquad (4)$$

can be set up, which is found to be of the form

$$\tilde{G}^a_b = G^a_b + \rho^a_b. \qquad (5)$$

What is truly remarkable about this result, as first discovered in [35, 36], is that the deformed part of the mixed Einstein tensor

$$\rho^a_{\ b} = -\frac{1}{2} f R^a_{\ c} l^c l_b - \frac{1}{2} f R^c_{\ b} l_c l^a + \frac{1}{2} \delta^a_b (f R^d_{\ c} l_d l^c - \nabla_d \nabla^c (f l^d l_c)) + \quad (6)$$
$$+ \frac{1}{2} (\nabla_c \nabla^a (f l^c l_b) + \nabla^c \nabla_b (f l_c l^a) - \nabla_c \nabla^c (f l^a l_b))$$

is linear in the profile function $f$.

Based on the fact that the considered Kerr-Schild vector field $l^a$ can always be completed to a normalized null geodesic frame $(l^a, k^a, m^a, \bar{m}^a)$ whose components satisfy $-k_a l^a = m_a \bar{m}^a = 1$, the spin-coefficient method of Newman and Penrose [30] can be applied to the given problem. An appropriate candidate for the corresponding Kerr-Schild vector field $l^a$ can be found by performing a $2+2$-decomposition of the given seed metric and its inverse, which leads to the expressions $g_{ab} = -2l_{(a} k_{b)} + 2m_{(a} \bar{m}_{b)}$ and $g^{ab} = -2l^{(a} k^{b)} + 2m^{(a} \bar{m}^{b)}$. This gives a null tetrad $(l^a, k^a, m^a, \bar{m}^a)$ that is of the form

$$l^a = \frac{r^2 + a^2}{\Sigma} \partial_v^a + \frac{a}{\Sigma} \partial_\phi^a + \frac{\Delta}{2\Sigma} \partial_r^a \quad (7)$$
$$k^a = -\partial_r^a$$
$$m^a = \frac{1}{\sqrt{2}\Gamma} (\partial_\theta^a - \frac{i}{\sin\theta} (\partial_\phi^a + a \sin^2\theta \partial_v^a))$$
$$\bar{m}^a = \frac{1}{\sqrt{2}\bar{\Gamma}} (\partial_\theta^a + \frac{i}{\sin\theta} (\partial_\phi^a + a \sin^2\theta \partial_v^a)),$$

where $\Gamma = r + ia\cos\theta$ and $\bar{\Gamma} = r - ia\cos\theta$.

In respect to this particular choice of a null frame, all the nonzero spin-coefficients can be calculated. These coefficients are

$$\epsilon = \frac{r - M}{2\Sigma} - \frac{\Delta}{2\Sigma\Gamma}, \ \alpha = -\frac{\cot\theta}{2\sqrt{2}\bar{\Gamma}}, \ \tau = \frac{ia\sin\theta}{\sqrt{2}\Gamma^2} \quad (8)$$
$$\beta = \frac{ia\sin\theta}{\sqrt{2}\Gamma^2} + \frac{\cot\theta}{2\sqrt{2}\Gamma}, \ \pi = -\frac{ia\sin\theta}{\sqrt{2}\Sigma}, \ \rho = -\frac{\Delta}{2\Sigma\Gamma}, \ \mu = -\frac{1}{\Gamma},$$
$$\phi_1 = \frac{e}{\sqrt{2}\Gamma^2}, \ \Psi_2 = -\frac{M}{\Gamma^3} + \frac{e^2}{\Gamma^2\Sigma}.$$

As can be seen, $\epsilon + \bar{\epsilon} \neq 0$ applies according to the given choice of the null frame. Hence, to actually convert the given expression into a null geodesic frame, a null rescaling of the form

$$l^a \to B e^{-\kappa v} l^a, \quad k^a \to B^{-1} e^{\kappa v} k^a \quad (9)$$

needs to be applied, where

$$B = B(r, \theta) = B_0 \Sigma \exp(-\frac{r}{M} - \frac{2M^2 - e^2}{2M(r_+^2 + a^2)} [r + 2M \ln|r - r_-|]) \quad (10)$$

with

$$B_0 = B_0(r,\theta) = \frac{\exp(\frac{r_+}{M} + \frac{2M^2-e^2}{2M(r_+^2+a^2)}[r_+ + 2M\ln|r_+ - r_-|])}{r_+^2 + a^2\cos^2\theta} \quad (11)$$

and $\kappa = \frac{r_+ - M}{r_+^2 + a^2}$. Here, as can readily be seen, the resulting expression remains regular at the outer event horizon, that is, for a radial parameter value of $r = r_+$.

By performing this rescaling, the condition $\epsilon + \bar{\epsilon} = 0$ is now exactly satisfied. As a consequence, using the definitions $D' := k^b \nabla_b$, $\delta := m^b \nabla_b$ and $\delta' := \bar{m}^b \nabla_b$, the structure of the associated spin-coefficients changes according to the rule

$$\epsilon \to \epsilon + \frac{1}{2}D\ln|Be^{-\kappa v}|,\ \gamma \to \gamma + \frac{1}{2}D'\ln|Be^{-\kappa v}|,\ \gamma \to \gamma + \frac{1}{2}D'\ln|Be^{-\kappa v}|,$$
$$\alpha \to \alpha + \frac{1}{2}\delta'\ln|Be^{-\kappa v}|\ \beta \to \beta + \frac{1}{2}\delta\ln|Be^{-\kappa v}|,$$
$$\rho \to Be^{-\kappa v}\rho,\ \mu \to B^{-1}e^{\kappa v}\mu; \quad (12)$$

all other coefficients are zero or remain unaltered.

Before explicitly calculating the deformed part $\rho_b^a$ of the total Einstein tensor $\tilde{G}_b^a$, it shall first be noted that the geometric field of a null particle field at the horizon must satisfy $\tilde{G}_b^a|_{\mathcal{H}^+} - G_b^a|_{\mathcal{H}^+} \propto l^a l_b$. However, since this obviously implies that $\tilde{R}|_{\mathcal{H}^+} = 0$ as a consequence of $R = 24\Pi = 0$, there must further hold $\nabla_d \nabla^c(f l^d l_c) = 0$ locally at the event horizon of the black hole background. This is certainly satisfied if the local condition

$$\nabla^c(f l^d l_c)|_{\mathcal{H}^+} = 0 \quad (13)$$

is met by the profile 'function' $f$ of the geometry.

To describe the distributional profile of a point-like null particle, it is appropriate to consider an ansatz that contains a delta distribution. More precisely, a generic choice for $f = f(v, r, \theta, \phi)$ that is consistent with the above requirements should locally be of the form

$$f(v,r,\theta,\phi)|_{\mathcal{H}^+} = e^{\kappa v} f_0(v,\theta,\phi)\delta(r-r_+) = U(\theta)F(\theta,\phi-\omega_+ v)e^{\kappa v}\delta(r-r_+), \quad (14)$$

where $\omega_+ = \frac{a}{r_+^2 + a^2}$ and $U = U(\theta)$ and $F(\theta, \phi - \omega_+ v)$ are free functions that are to be specified in the further course of this work.

Since the deformed Einstein tensor decomposes according to the rule (5), where $G_b^a$ is the Einstein tensor Kerr-Newman background geometry, it can be concluded from the fact that the background field equations (Einstein-Maxwell equations) are satisfied that the total stress-energy tensor of the deformed spacetime must be of the form

$$\tilde{T}_b^a = T_b^a + \tau_b^a, \quad (15)$$

where $T_b^a = F^{ac}F_{cb} - \frac{1}{4}\delta_b^a F_{cd}F^{cd}$ is the electromagnetic stress-energy tensor of the Kerr-Newman background geometry and $\tau_b^a = \tilde{\Phi}_{22} l^a l_b = \varepsilon_0 \delta_+ \delta_N l^a l_b$ with $\delta_+ := \delta(r - r_+)$ and $\delta_N := \delta(\cos\theta - 1)$ is the energy-momentum tensor of the massless particle. As usual, the electromagnetic part of the stress-energy tensor is written down in terms of the Faraday tensor $F_{ab} = 2\nabla_{[a}A_{b]}$, i.e. the exterior derivative of the one-form potential $A_b$.

Consequently, considering the fact that the Einstein equations decouple completely in background and particle field equations, one can see that no distortion of the background field of the charged black hole occurs through interaction with the particle field. Instead, the Einstein equations can be split into two different parts that can be solved separately, where, of course, the Kerr-Schild part is not independent of the choice of geometric background, but inherits essential properties of the Kerr-Newman spacetime in relation to which it is defined. The deformed mixed energy-momentum tensor $\tau^a_b$ of this part of the field equations is that of a null fluid with a null energy density proportional to a product of the Dirac delta distributions.

Consequently, however, one finds that it suffices to solve

$$\rho^a_b = 8\pi \tau^a_b \qquad (16)$$

on $\mathcal{H}^+$ and therefore to calculate the local object

$$\rho^a_b = [G^a_b] = \tilde{G}^a_b|_{\mathcal{H}^+} - G^a_b|_{\mathcal{H}^+} \qquad (17)$$

in order to determine the geometric structure of the deformed spacetime $(\tilde{M}, \tilde{g})$. This greatly simplifies a rather lengthy calculation, which shows that the left side of the equations reduces to a single relation of the form

$$\rho^a_b = -\boxtimes f l^a l_b, \qquad (18)$$

where

$$\boxtimes := \frac{1}{2}\mathcal{D}^2 + (2(\alpha + \bar{\beta}) - \bar{\tau})\delta + (2(\bar{\alpha} + \beta) - \tau)\delta' + \frac{1}{2}D'(\rho + \bar{\rho}) + \qquad (19)$$
$$+ \delta(\alpha + \bar{\beta} - \bar{\tau}) + \delta'(\bar{\alpha} + \beta - \tau) + (\bar{\tau} + \alpha - \bar{\beta})(\tau - \bar{\alpha} - \beta) +$$
$$+ (\tau + \bar{\alpha} - \beta)(\bar{\tau} - \alpha - \bar{\beta}) + 4|\alpha + \bar{\beta}| - \bar{\tau}(\bar{\alpha} + \beta) +$$
$$- \tau(\alpha + \bar{\beta}) + \Psi_2 + \bar{\Psi}_2.$$

Note that the relations $\frac{1}{2}[(\rho + \bar{\rho})D'f, \cdot] = -\frac{1}{2}[D'(\rho + \bar{\rho})f, \cdot]$ and $-\frac{1}{2}[R^a_c l^c l_b + R^c_b l^a l_c] = -2\Phi_{11} l^a l_b = -2|\phi_1|^2 l^a l_b$ have been used in this context to determine the concrete form of this differential operator.

Unfortunately, by looking at the result obtained, it immediately becomes clear that the derived differential relation is too complicated to be solved directly in its present form. Consequently, in order to further simplify the geometric structure of the Kerr-Schild deformed Einstein tensor, it proves beneficial to try to satisfy the local conditions

$$\nabla_{[a} l_{b]}|_{\mathcal{H}^+} = 0, \qquad (20)$$

which can be satisfied by performing a null rotation

$$l^a \to l^a,\ k^a \to k^a + \bar{\zeta}m^a + \zeta\bar{m}^a + |\zeta|l^a,\ m^a \to m^a + \zeta l^a, \tag{21}$$

where $\zeta = \frac{i\omega_+ e^{\kappa v}\bar{\Gamma}\sin\theta}{\sqrt{2}B}$ shall apply by definition. In the new null frame, the spin-coefficients as well as the Weyl and Maxwell scalars take a different form. How exactly these objects change in the process can be read in [13, 27]. The non-zero Weyl and Maxwell scalars are $\Psi_2 \to \Psi_2\ \Psi_3 \to 3\bar{\zeta}\Psi_2$, $\Psi_4 \to 6\bar{\zeta}^2\Psi_2$ and $\phi_1 \to \phi_1$ and $\phi_2 \to \phi_2 + 2\bar{\zeta}\phi_1$. Although these changes appear to alter the structure of the deformed field equations (16) considerably, relation (6), as it turns out, is invariant under the given null transformation. What changes, however, is the form of the spin-coefficients and the structure of the complex-valued null normals $m^a$ and $\bar{m}^a$, which changes in such a way that $\bar{\alpha} + \beta = \tau$ is satisfied on $\mathcal{H}^+$ and $\nabla_{[a}l_{b]}|_{\mathcal{H}^+} = 0$ holds as a further consequence. Consequently, however, it is found that the deformed Einstein tensor can be written exactly the same way as in equation (18), with the only difference being that the scalar part of the said relation now reads

$$\boxtimes f = \frac{1}{2}\mathcal{D}^2 f + \tau\delta'f + \bar{\tau}\delta f + \frac{1}{2}D'(\rho+\bar{\rho})f + 2|\tau|^2 f + \Psi_2 f + \bar{\Psi}_2 f, \tag{22}$$

provided that the definition $\mathcal{D}^2 := \mathcal{D}^a\mathcal{D}_a = \delta\delta' + \delta'\delta + (\beta-\bar{\alpha})\delta' + (\bar{\beta}-\alpha)\delta$ is used in the present context.

Remarkably, the given relation (22), which was previously found in respect to other definitions and conventions in [10], is invariant under null rotations that leave the form of $l^a$ invariant. Accordingly, the result obtained remains completely unchanged if a corresponding null rotation is considered in the present context, which, however, ensures that complex null normals $m^a$ and $\bar{m}^a$ are hypersurface forming, so that $L_{\bar{m}}m^a$ is given by a linear combination in $m^a$ and $\bar{m}^a$ locally on $\mathcal{H}^+$. Consequently, it seems justified to continue using the null reference frame considered so far.

Although not obvious at first glance, the differential operator given in expression (22) is already known in the literature. To see this, one may consider the spin-coefficient relations [30, 31]

$$D'\rho - \delta'\tau + |\tau|^2 - \tau(\bar{\beta}-\alpha) + \Psi_2 = 0 \tag{23}$$

and

$$\delta'\beta - \delta\alpha + |\alpha|^2 + |\beta|^2 - 2\alpha\beta - \Psi_2 + \Phi_{11} = 0 \tag{24}$$

and their respective complex conjugates and use them to convert (16). For this purpose, one may additionally use the definition of the Gauß curvature $K = \delta(\alpha-\bar{\beta}) + \delta'(\bar{\alpha}-\beta) - 2(\alpha-\bar{\beta})(\bar{\alpha}-\beta)$ associated with the induced metric $q_{ab} = m_a\bar{m}_b + \bar{m}_am_b$ in oder to obtain the result

$$\boxtimes f = \frac{1}{2}\left[\mathcal{D}^2 + 2\tau\delta' + 2\bar{\tau}\delta\right]f+ \tag{25}$$
$$+ \frac{1}{2}\left[\delta'\tau + \delta\bar{\tau} + \tau(\bar{\beta}-\alpha) + \bar{\tau}(\beta-\bar{\alpha}) + 2|\tau|^2 - K + 2\Phi_{11}\right]f,$$

which, by using the Hájiček one-form $\omega_a = (\alpha+\bar{\beta})m_a + (\bar{\alpha}+\beta)\bar{m}_a = \bar{\tau}m_a + \tau\bar{m}_a$ can be more compactly be written in the form

$$\boxtimes f = \frac{1}{2}\left[\mathcal{D}^2 + 2(\omega\mathcal{D}) + \mathcal{D}_a\omega^a + \omega^2 - \frac{1}{2}\mathcal{R} + \frac{1}{2}q^{ab}R_{ab}\right]f. \tag{26}$$

Note that $\mathcal{R}$ is the scalar curvature of a section $\mathcal{S}$ of the exterior event horizon $\mathcal{H}^+$.

One may recall that the profile function of the geometry has previously been chosen to be of the form $f|_{\mathcal{H}^+} = U(\theta)F(\theta, \phi - \omega_+ v)e^{\kappa v}\delta(r - r_+)$ on the exterior event horizon of the black hole. Consequently, making an ansatz of the form $U = U(\theta)$, where $U$ is a solution to the equation

$$\delta \ln|U| = \bar{\pi} - \tau \tag{27}$$

on $\mathcal{H}^+$, and assuming that $F = F(\theta)$, it follows that relation (22) takes the form

$$\boxtimes f = e^{\kappa v} U \delta_+ (\frac{1}{2}\mathcal{D}^2 F + \pi\delta F + \bar{\pi}\delta' F + \frac{1}{2}D'(\rho + \bar{\rho})F -$$
$$- \frac{1}{2}\delta\bar{\tau}F - \frac{1}{2}\delta'\tau F + |\pi|^2 F + |\tau|^2 F + \frac{1}{2}(\bar{\pi} - \tau)(\bar{\beta} - \alpha)F +$$
$$+ \frac{1}{2}(\pi - \bar{\tau})(\beta - \bar{\alpha})F + \Psi_2 F + \bar{\Psi}_2 F), \tag{28}$$

which looks quite complicated at first sight. However, by using relation (23) and taking into account that

$$\delta\pi - D\mu + |\pi|^2 + \pi(\bar{\beta} - \alpha) + \Psi_2 = 0, \tag{29}$$

one finds that the geometric structure of the deformed Einstein tensor drastically simplifies in that the equation for the profile function now takes the much simpler form

$$\boxtimes f = \frac{e^{\kappa v}\delta_+ U}{2} \boxplus F. \tag{30}$$

where $\boxplus \equiv \mathcal{D}^2 + D(\mu + \bar{\mu}) = \Delta - 2(\omega\mathcal{D}) + D(\mu + \bar{\mu})$. Note that the identity $\bar{\tau}\delta F + \tau\delta' F = -\pi\delta F - \bar{\pi}\delta' F$, which is specific to the given case of Kerr-Newman spacetime and does not apply in the general case, has been used to obtain this form of the differential operator $\boxplus$.

It is worth mentioning in this context that (with a little care) relations (22) and (30) can also be obtained in a different way, namely by using not the field equations of the theory, but only the Newman-Penrose spin-coefficient formalism. More specifically, by considering the $2 + 2$-decompositions $\tilde{g}_{ab} = -2l_{(a}\tilde{k}_{b)} + 2m_{(a}\bar{m}_{b)}$ and $\tilde{g}^{ab} = -2l^{(a}\tilde{k}^{b)} + 2m^{(a}\bar{m}^{b)}$ of the Kerr-Schild deformed metric and its inverse, where $\tilde{k}_a := k_a - \frac{f}{2}l_a$ and $\tilde{k}^a := k^a + \frac{f}{2}l^a$, the spin-coefficient formalism can be used to derive relations between the spin-coefficients of the Kerr-Newman background spacetime $(M, g)$ and its Kerr-Schild deformed counterpart $(\tilde{M}, \tilde{g})$. These relations have been obtained by Bilge and Gürses in [12]. A specific spin-coefficient relation[1] that proves to be relevant in this context is the following:

$$\frac{1}{2}[\delta\tilde{\nu} + (2\tilde{\beta} - \tilde{\bar{\pi}})\nu + \delta\bar{\tilde{\nu}} + (2\bar{\tilde{\beta}} - \tilde{\pi})\bar{\tilde{\nu}}+ \tag{31}$$
$$-\tilde{D}'(\tilde{\mu}+\bar{\tilde{\mu}}) + (\tilde{\gamma}+\bar{\tilde{\gamma}})(\tilde{\mu}+\bar{\tilde{\mu}}) + \tilde{\mu}^2 + \bar{\tilde{\mu}}^2] = \tilde{\Phi}_{22},$$

with $\tilde{\mu} = \mu + \rho f$, $\tilde{\nu} = \delta' f + (\bar{\tau}-\pi)f$ and $\tilde{\beta} = \beta$, $\tilde{\pi} = \pi$ and $\tilde{\gamma}+\bar{\tilde{\gamma}} = \gamma+\bar{\gamma}$. Due to the fact that $f(v,r,\theta,\phi) \propto \delta(r-r_+)$, where $\delta(r-r_+)$ is Dirac's delta distribution, this relation looks mathematically ill-defined and therefore pathological, because it seems to contain undefined 'squares' of the delta distribution. However, by considering a delta sequence $\delta_\varepsilon$ which coincides with the delta distribution in the limit $\varepsilon \to 0$ instead of naively trying to work directly with the delta distribution, thereby using methods from Colombeau's theory of generalized functions [14, 15, 19], one finds that this is actually not the case. That is to say, a careful analysis shows in this context that $\tilde{\mu} \approx \mu$ holds on the exterior black hole event horizon $\mathcal{H}^+$ and that also $\tilde{\mu}^2 \approx \mu^2$ is locally fulfilled on the same hypersurface, where $\approx$ means association in the sense of distributions. Moreover, one finds that $\tilde{D}'(\tilde{\mu}+\bar{\tilde{\mu}}) \approx D'(\mu+\bar{\mu})+\frac{f}{2}D(\mu+\bar{\mu})$ holds on the black hole horizon. Consequently, however, it turns out that equations (31) is actually well-defined and coincides exactly with relation (22), which confirms the results found in [21]. To see this, one may take into account that $D(\mu+\bar{\mu}) = -D'(\rho+\bar{\rho})$ on $\mathcal{H}^+$ and, moreover, the fact that the analog of equation (31) in the background field is $-D'(\mu+\bar{\mu})+\mu^2+\bar{\mu}^2+(\gamma+\bar{\gamma})(\mu+\bar{\mu}) = 0$. Using this and relations (23) and (29), one finds that the left hand side of (31) coincides exactly with (22). Consequently, making again an ansatz of the form $f|_{\mathcal{H}^+} = U(\theta)F(\theta,\phi-\omega_+ v)e^{\kappa v}\delta(r-r_+)$, where $U(\theta)$ is a solution to the differential equation $\delta U + (\tau-\bar{\pi})U = 0$, equation (30) results once again, but this time from (31). It may be noted the same result can also be determined by using the findings of [12].

Although usually defined with respect to a different choice for the null tetrad $(l^a, k^a, m^a, \bar{m}^a)$, the elliptic differential operator $\boxtimes$ appearing in $(25-26)$, (28) and (30) is actually also well known in the literature on black hole physics, that is, from the theory of isolated horizons of Ashtekar et al. [4, 5, 6, 7, 8, 9], where its negative is used to determine the genericity of weakly isolated horizons (WIH) (see page 17 of [7]). A relation similar to (30) should therefore also exist in that particular approach and may even prove useful in finding the eigenfunctions of the associated differential operator.

Interestingly, it also turns out that the differential operator $\boxtimes$ is closely related to another differential operator, namely a stability operator for marginally outer trapped surfaces (MOTS) presented in [3]. To see this, consider eigenvalue

---

[1] Note that this specific form of the relation results from adding $(a')$ on page 248 of [30] and its complex conjugate.

problem

$$L_\mathcal{S}\psi = \left[-\Delta + 2(\omega\mathcal{D}) + \mathcal{D}_a\omega^a - \omega^2 + \frac{1}{2}\mathcal{R} - G_{ab}l^ak^b\right]\psi \qquad (32)$$
$$= -\left[(\mathcal{D}-\omega)^2 - \frac{1}{2}\mathcal{R} + G_{ab}l^ak^b\right]\psi = \lambda\psi$$

which forms the core of the the stability analysis for MOTS. Evidently, in the case that $\omega^a = 0$ holds, it can be concluded that $\boxtimes \equiv -\frac{1}{2}L_\mathcal{S}$. Thus, for the special cases of Reissner-Nordström and Schwarzschild spacetimes, both differential operators agree up to a constant factor. In addition, however, there is yet another interesting connection between these two differential operators. To see this, one may follow the results of [23, 24], where a fascinating mathematical analogy between the MOTS stability operator and the Hamiltonian of a charged particle was provided by making the identifcations $\omega_a \to \frac{ie}{\hbar c}A_a$, $\mathcal{R} \to \frac{4me}{\hbar^2}\phi$, $G_{ab}l^ak^b \to -\frac{2m}{\hbar}V$ in relation (32). For through these identifications, said relation becomes a Schrödinger equation for a non-relativistic charged particle moving in an electromagnetic field with magnetic vector potential $A_a$ and electric potential $\phi$, and the MOTS stability operator takes the exact form of an associated Hamiltonian $H \longleftrightarrow \frac{\hbar^2}{2m}L_\mathcal{S}$, so that

$$H\psi = \left[-\frac{1}{2m}(i\hbar\mathcal{D} - \frac{e}{c}A)^2 + e\phi + V\right]\psi. \qquad (33)$$

Interestingly, making the similar identifications $\omega_a \to -\frac{ie}{\hbar c}A_a$, $\mathcal{R} \to \frac{4me}{\hbar^2}\phi$, $G_{ab}l^ak^b \to -\frac{2m}{\hbar}V$, $-\frac{f}{2} \to \psi$ in (27), one obtains the exact same relation (33), which, however, implies that also (27) has striking similarities to a Schrödinger equation for a charged particle, although both the MOTS stability operator $L_\mathcal{S}$ and the shock wave operator $\boxtimes$ appear to be generically non-selfadjoint, in stark contrast to the Hamiltonian $H$.

Nevertheless, as far as the shock wave operator $\boxtimes$ is concerned, this turns out not to be the case. The reason for this is that by choosing appropriate conventions, which will be introduced in the next section, relation (23) can be re-written in the form

$$\boxtimes f = \Sigma_+^{-1}(\Delta_{\mathbb{S}_2} + V)f, \qquad (34)$$

where $V = V(\cos\theta)$ is a potential term, which immediately makes it clear that $\boxtimes$ is self-adjoint.

This is of interest because, although it is generally expected that the MOTS stability operator is not self-adjoint, it can be checked by direct calculation that said operator is actually self-adjoint in the given Kerr-Newman case the same way as the elliptic shock wave operator $\boxtimes$. Consequently, taking the fact into account that the geometric structure needed to set up relation (32) is similar for different spacetime geometries, it can plausibly be conjectured that the self-adjointness of $L_\mathcal{S}$ can also be proven for classes of spacetimes more general than the Kerr-Newman class such as, for instance, for the class of rotating Bonnor-Vaidya spacetimes.

Anyway, the self-adjointness of $L_S$ proves to be of interest in the given case of Kerr-Newman spacetime not least because there is yet another common characteristic between the MOTS stability approach and the quantum particle approach, which is also common to the null particle shock wave approach discussed so far, but which proves to be somewhat more subtle in the latter case. This common characteristic is that both equations (32) and (33) are invariant under the combined transformations $\omega_a \to \omega_a - \mathcal{D}_a\sigma, \psi \to e^{-\sigma}\psi$ and $A_a \to A_a - \mathcal{D}_a\sigma$, $\psi \to e^{\frac{ie}{\hbar c}\sigma}\psi$ and therefore define Abelian gauge symmetries. To be more specific, the QCP Schrödinger equation (33) has an Abelian $U(1)$-gauge symmetry, whereas the MOTS stability relation (32) has an Abelian gauge symmetry resulting from the gauge freedom in the definition of the rotation one-form $\omega_a$, which occurs due to the fact that the normalization of the future-directed null vectors $l^a$ and $k^a$ leaves a rescaling freedom by a positive function $e^{-\sigma}$, corresponding to a boost transformation $l^a \to e^{-\sigma}l^a$ and $k^a \to e^{\sigma}k^a$.

As it turns out, the shock wave case is a little more complicated. This is because a boost transformation of the form $l^a \to e^{-\sigma}l^a$ and $k^a \to e^{\sigma}k^a$ completely changes the physical character of the gravitational shock wave field by leading to a different deformation tensor $\tilde{\rho}^a_b = -\tilde{\boxtimes}\tilde{f}\tilde{l}^a\tilde{l}_b = -\tilde{\boxtimes}\tilde{f}e^{-2\sigma}l^al_b$. Nevertheless, as can be concluded from the concrete form of (6), there is an Abelian gauge symmetry also in the case of a gravitational shock wave on the event horizon of a Kerr-Newman black hole, which becomes apparent by taking into account that the boost transformation mentioned before leaves the deformed Einstein tensor $\rho^a_b$ invariant if (and only if) the profile function of the geometry "scales" according to the rule $f \to e^{2\sigma}f$.

From this, however, it can be concluded that the transformation mentioned has no effect on the field equations of the theory if $\omega_a \to \omega_a - \mathcal{D}_a\sigma, f \to e^{2\sigma}f$, which further implies that said field equations can again be written down in terms of the shock wave operator $\boxtimes$ defined in (26). Therefore, one comes to the conclusion that field equation (26), which describes the field of a gravitational shock wave caused by a null particle on the horizon of a Kerr-Newman black hole, has similar properties as the Schrödinger equation of a quantum charged particle (33) and the MOTS stability equation (32) in certain respects.

Anyhow, using the results (16), (18) and (22), one obtains the scalar field equation
$$\boxtimes f = \Sigma_+^{-1}(\Delta_{\mathbb{S}_2} + V)f = 2\pi d_0 \delta_N, \tag{35}$$
which shall be referred to as generalized Dray-'t Hooft equation from now on.

By considering the limit $a \to 0$, this relation takes the form
$$(\Delta_{\mathbb{S}_2} - c)f = 2\pi b_0 \delta_N, \tag{36}$$
where $c = 2\kappa r_+ = \frac{2(r_+ - M)}{r_+}$, which was first obtained by Sfetsos. In addition, by considering the combined limits $a, e \to 0$, said differential equation takes the even simpler form
$$(\Delta_{\mathbb{S}_2} - 1)f = 2\pi b_0 \delta_N, \tag{37}$$

which was first found by Dray and 't Hooft. Therefore, since both of these relations result as a special case from the given model, it can be concluded that the generalized Dray-'t Hooft equation provides a viable extension of the fundamental equations, which determine the profile functions corresponding to a gravitational shock wave in either Reissner-Nordström or Schwarzschild black hole backgrounds.

The goal of the next two sections will be to show that the generalized Dray-'t Hooft equation (35) can indeed be solved in a distributional sense, to discuss the main properties of the solution obtained and to show that the solution reduces to the solutions of Dray and 't Hooft and Sfetsos in a suitable limit.

## 2 Einstein's Field Equations II: Solving the generalized Dray-'t Hooft Relation

Next, to actually solve the deduced differential relation (35), the line element associated with the induced two-metric of the Kerr-Newman geometry

$$d\sigma^2 = \Sigma_+ d\theta^2 + \frac{(r_+^2 + a^2)^2 \sin^2\theta}{\Sigma_+} d\phi^2 \tag{38}$$

shall be considered. By introducing the coordinate transformation $\xi = \cos\theta$, this line element can be re-written in the form

$$d\sigma^2 = \frac{\Sigma_+}{1-\xi^2} d\xi^2 + \frac{(r_+^2 + a^2)^2 (1-\xi^2)}{\Sigma_+} d\phi^2, \tag{39}$$

which provides a spacelike dyad $(m^a, \bar{m}^a)$ that is locally of the form

$$m^a = \frac{1}{\sqrt{2}\Gamma_+}(\sqrt{1-\xi^2}\partial_\xi^a - \frac{i\Sigma_+}{(r_+^2+a^2)\sqrt{1-\xi^2}}\partial_\phi^a), \tag{40}$$

$$\bar{m}^a = \frac{1}{\sqrt{2}\bar{\Gamma}_+}(\sqrt{1-\xi^2}\partial_\xi^a + \frac{i\Sigma_+}{(r_+^2+a^2)\sqrt{1-\xi^2}}\partial_\phi^a), \tag{41}$$

according to which, of course, $\Gamma_+ = r_+ + ia\xi$ and thus $\bar{\Gamma}_+ = r_+ - ia\xi$.

Given this setting, the reduced generalized Dray-'t Hooft equation may be re-written in these coordinates via using (30) with $U = \Sigma_+^{-1}$ in the form

$$\Sigma_+^{-1}(\mathcal{D}^2 F - 2(\omega \mathcal{D})F + V \cdot F) = \Sigma_+^{-1}[\partial_\xi(\frac{1-\xi^2}{\Sigma_+}\partial_\xi)F - \tag{42}$$

$$+\frac{\Sigma_+^2}{(r_+^2+a^2)^2(1-\xi^2)}\partial_\phi^2 F - \frac{2r_+(r_+ - M)}{\Sigma_+}F].$$

By considering in the following the special case of a profile function $f = f(\xi)$ as well as the reduced profile function $F = F(\xi)$ which both do not depend on the angular variable $\phi$, while assuming that the symmetry axis of the system points through the 'north pole', i.e. through the point $\xi_0 = 1$ at which the particle is assumed to be located, it is possible to re-write the left hand side of relation (22) in the form

$$\boxtimes f = \frac{e^{\kappa v}\delta_+}{\Sigma_+} \boxplus F = \qquad (43)$$
$$= \frac{e^{\kappa v}\delta_+}{\Sigma_+} \left\{ \frac{d}{d\xi}\left(\frac{1-\xi^2}{\Sigma_+}\frac{dF}{d\xi}\right) - \frac{2r_+(r_+ - M)}{\Sigma_+}F \right\}$$

Note that the same step was also taken in the previous works of Dray and 't Hooft and Sfetsos, although, as one must admit, in the spherical case (quite contrary to the given axisymmetric case) such an approach does not result in any loss of generality.

The first step in finding a distributionally defined solution of the generalized Dray-'t Hooft relation

$$[\boxtimes f, \cdot] = [2\pi b_0 \delta(\xi - 1), \cdot] \qquad (44)$$

is to solve $\boxtimes f = 0$.

Relation (43) yields the homogeneous linear ordinary differential relation

$$\frac{d^2 F}{d\xi^2} + p\frac{dF}{d\xi} + qF = 0, \qquad (45)$$

where the coefficients $a$ and $b$ are given by $p = -\frac{2\xi}{1-\xi^2} - \frac{2a^2\xi}{\Sigma_+}$ and $q = -\frac{2r_+(r_+ - M)}{\Sigma_+(1-\xi^2)}$. Evidently, the coefficients give rise to the five singular points $\pm 1$, $\pm\frac{ir_+}{a}$ and $\infty$. Fortunately, however, since one finds in this context that the expressions $\lim_{\xi \to \xi_0}(\xi - \xi_0)p$ and $\lim_{\xi \to \xi_0}(\xi - \xi_0)^2 q$ remain finite, where $\xi_0$ represents any of the listed singularities, the given singular points happen to be regular singular points, which makes it clear that the derived differential equation belongs to the Fuchsian class of linear differential equations of second order with regular singular coefficients [16, 33].

In fact, from this it can be concluded that there must exist two different solutions that can be superimposed to a single solution. To determine the exact form of these solutions on the level of infinite power series, the given relation shall be re-written in the form

$$\Sigma_+(1-\xi^2)\frac{d^2F}{d\xi^2} - [2\xi\Sigma_+ + 2a^2\xi(1-\xi^2)]\frac{dF}{d\xi} - 2r_+(r_+ - M)F = 0. \qquad (46)$$

By doing so, it is found that the two solutions are of the form $F_1 = \sum_{k=0}^{\infty} w_k \xi^k$ and $F_2 \equiv F_1 \Delta + G$, where $G = \sum_{k=0}^{\infty} u_k \xi^k$ and $\Delta = C_1[\frac{1}{2}\ln|\frac{\xi+1}{\xi-1}| - a\omega_+\xi] + C_2$, where, in this context, $\Delta$ is then defined in precisely such a way that it forms a solution of

$$\Sigma_+(1-\xi^2)\frac{d^2\Delta}{d\xi^2} - [2\xi\Sigma_+ + 2a^2\xi(1-\xi^2)]\frac{d\Delta}{d\xi} = 0. \qquad (47)$$

Unfortunately, however, the derived differential relation (in its present form) does not allow for any polynomial solutions so that both $F_1 = F_1(\xi)$ and $F_2 = F_2(\xi)$ are in fact infinite power series. While this does not imply that there could not in principle exist another variable more appropriate than $\xi$, which allows for the definition of polynomial solutions and could therefore be used as a basis for the construction of generalized ellipsoidal harmonics, the finding that there are no such solutions in the given variable $\xi$ nevertheless shows that another pair of solutions is needed at this stage in order to find a combined solution not only to the homogeneous, but also the inhomogeneous equation.

The independent solutions $F_1^\pm = F_1^\pm(\xi)$ and $F_2^\pm = F_2^\pm(\xi)$, which can be obtained straightforwardly by introducing a new set of variables $\xi \to \pm\xi - 1$, are of the form $F_1^\pm = \sum_{k=0}^\infty (-1)^k w_k^\pm (1 \mp \xi)^k$ and $F_2^\pm \equiv F_1^\pm \Delta + G^\pm$, where $G^\pm = \sum_{k=0}^\infty (-1)^k u_k^\pm (1 \mp \xi)^k$. As it then turns out, the corresponding coefficients must coincide exactly in that $w_k^+ = w_k^-$ and $u_k^+ = u_k^-$ as long as the conditions $w_0^+ = w_0^-$ and $u_0^+ = u_0^-$ are satisfied, which is simply due to the invariance of the equation under 'parity transformations' in $\xi$. The coefficients $w_k^\pm$ have a rather complicated structure, as they have to be determined from the three-term recurrence relation

$$w_{k+3}^\pm = \sum_{j=1}^3 m_{k,j} w_{k+3-j}^\pm, \tag{48}$$

according to which one has $m_{k,1} = -\frac{r_+^2}{2(r_+^2+a^2)}\frac{k(k+1)}{(k+1)^2} - \frac{a^2}{2(r_+^2+a^2)}\frac{(5k-3)k}{(k+1)^2} - \frac{\kappa r_+}{(k+1)^2}$, $m_{k,2} = -\frac{2a^2}{r_+^2+a^2}\frac{(k-2)(k-1)}{(k+1)^2}$ and $m_{k,3} = -\frac{a^2}{r_+^2+a^2}\frac{(k-2)(k-3)}{(k+1)^2}$.

This recurrence relation can be solved by using Mallik's companion matrix approach [28]. This method is based on the fact that relation (39) can be expressed in vector form as

$$\begin{pmatrix} w_{k+3}^\pm \\ w_{k+2}^\pm \\ w_{k+1}^\pm \end{pmatrix} = \begin{pmatrix} m_{k,1} & m_{k,2} & m_{k,3} \\ 1 & 0 & 0 \\ 0 & 1 & 0 \end{pmatrix} \begin{pmatrix} w_{k+2}^\pm \\ w_{k+1}^\pm \\ w_k^\pm \end{pmatrix}, \tag{49}$$

where the $3 \times 3$ matrix on the right hand side of (49) is the so-called companion matrix for index $k$. Defining $W_{k+1}^\pm := \left(w_{k+2}^\pm, w_{k+1}^\pm, w_k^\pm\right)^\mathsf{T}$ and $M_k := \begin{pmatrix} m_{k,1} & m_{k,2} & m_{k,3} \\ 1 & 0 & 0 \\ 0 & 1 & 0 \end{pmatrix}$, this relation can be re-written in the form

$$W_{k+1}^\pm = M_k W_k^\pm. \tag{50}$$

Using the further definition $W_1^\pm := \left(w_3^\pm, w_2^\pm, w_1^\pm\right)^\mathsf{T}$, one then finds by continuous iteration

$$W_{k+1}^\pm = M_k M_{k-1}...M_1 W_1^\pm, \tag{51}$$

where for $k \geq 1$ the resulting product matrix $A_k := \prod\limits_{i=3}^{k} M_i$ is given by

$$A_k = \begin{pmatrix} a_{k,1} & a_{k,2} & a_{k,3} \\ a_{k-1,1} & a_{k-1,2} & a_{k-1,3} \\ a_{k-2,1} & a_{k-2,2} & a_{k-2,3} \end{pmatrix} \qquad (52)$$

with components of the form

$$a_{k-i+1,j} = \sum_{r=1}^{k-i+j} \sum_{\substack{(l_1,l_2,...,l_r) \\ 1 \leq l_1,...,l_r \leq 3 \\ l_r \geq j \\ l_1+l_2+....+l_r=k-i+j}} \left[ \prod_{m=1}^{r} m_{k-i+1-\sum\limits_{n=1}^{m-1} l_n, l_m} \right] \qquad (53)$$

which are one if $i = j, i > k, k < 3$ and zero if $i \neq j, i > k, k < 3$.

Using then the fact that one can always choose w.l.o.g $w_2^{\pm}$ and $w_1^{\pm}$ freely and thus in such a way that $w_2^{\pm} = m_{1,1} w_1^{\pm} + m_{1,2} w_0^{\pm}$ and $w_1^{\pm} = m_{0,1} w_0^{\pm}$, so that $W_1^{\pm} = \left( w_2^{\pm}, w_1^{\pm}, w_0^{\pm} \right)^{\intercal} = \left( m_{1,1} w_1^{\pm} + m_{1,2} w_0^{\pm}, m_{0,1} w_0^{\pm}, w_0^{\pm} \right)^{\intercal}$ is met, relation (51) can be re-written in the form

$$W_{k+1}^{\pm} = M_k M_{k-1} ... M_0 W_0^{\pm}, \qquad (54)$$

where the definitions $M_0 = \begin{pmatrix} m_{1,1} & m_{1,2} & 0 \\ 1 & 0 & 0 \\ 0 & 1 & 0 \end{pmatrix} \begin{pmatrix} m_{0,1} & 0 & 0 \\ 1 & 0 & 0 \\ 0 & 1 & 0 \end{pmatrix}$ and $W_0^{\pm} = \left( w_0^{\pm}, w_0^{\pm}, w_0^{\pm} \right)^{\intercal}$ have been used.

Given these results, the form of the coefficient $w_{k+3}^{\pm}$ can straightforwardly be determined from (53) for all $k$, which allows one to conclude that $F_1^{\pm}$ actually represents a solution of differential equation (46).

Subsequently, taking into account the mathematical framework of the Fuchsian class [33], one immediately comes to the conclusion that the corresponding power series converges absolutely in the annulus of radius $|\xi - \xi_0| < R$.

The additional coefficients $u_k^{\pm}$ occurring in the definition of $G^{\pm}$ have a related, but a bit more complicated structure. They result from the inhomogeneous differential equation

$$\Sigma_+ (1-\xi^2) \frac{d^2 G^{\pm}}{d\xi^2} - [2\xi \Sigma_+ + 2a^2 \xi (1-\xi^2)] \frac{dG^{\pm}}{d\xi} - 2r_+ (r_+ - M) G^{\pm} = \qquad (55)$$

$$= -2\Sigma_+^2 \frac{dF_1^{\pm}}{d\xi}$$

for $G^{\pm}$, which results directly from inserting the second solution $F_2 = F_2(\xi)$ in the homogeneous differential equation (46).

Recalling here now the fact that $G^{\pm} = \sum\limits_{k=0}^{\infty} (-1)^k u_k^{\pm} (\xi \pm 1)^k$ applies and that it is always possible to set $2C_1 \Sigma_+^2 \frac{dF_1^{\pm}}{d\xi} = \sum\limits_{k=0}^{\infty} (-1)^k \phi_k^{\pm} (1 \pm \xi)^k$, where, using the definition $w_{k-j}^{\pm} = 0$ for $k < j$, the corresponding coefficients take the form $\phi_k^{\pm} := \sum_{j=0}^{4} \alpha_j(k-j+1)w_{k-j}^{\pm}$ with $\alpha_0 := 2C_1(r_+^2 + a^2)^2$, $\alpha_1 := 8C_1a^2(r_+^2 + a^2)$, $\alpha_2 := 4C_1a^2(r_+^2 + 3a^2)$, $\alpha_3 := 8C_1a^4$, $\alpha_4 := 4C_1a^4$, one finds the recurrence relation

$$u_{k+3}^{\pm} = \sum_{j=1}^{3} m_{k,j} u_{k+3-j}^{\pm} + \Phi_{k+3}^{\pm}, \tag{56}$$

which contains the expression $\Phi_k^{\pm} := \frac{\phi_k^{\pm}}{2(r_+^2+a^2)(k+3)^2}$. Defining $u_{k+3-j}^{\pm} = 0$ for $k+3 < j$, the coefficients $u_0^{\pm}$, $u_1^{\pm}$ and $u_2^{\pm}$ can straightforwardly be determined from (56). To obtain the remaining coefficients, one may use the fact that also relation (56) can be expressed in vector form, i.e.

$$\begin{pmatrix} u_{k+3}^{\pm} \\ u_{k+2}^{\pm} \\ u_{k+1}^{\pm} \end{pmatrix} = \begin{pmatrix} m_{k,1} & m_{k,2} & m_{k,3} \\ 1 & 0 & 0 \\ 0 & 1 & 0 \end{pmatrix} \begin{pmatrix} u_{k+2}^{\pm} \\ u_{k+1}^{\pm} \\ u_k^{\pm} \end{pmatrix} + \begin{pmatrix} \Phi_k^{\pm} \\ 0 \\ 0 \end{pmatrix}. \tag{57}$$

Defining $U_{k+1}^{\pm} := \left(u_{k+3}^{\pm}, u_{k+2}^{\pm}, u_{k+1}^{\pm}\right)^{\mathsf{T}}$ and $\Phi_k^{\pm} := \left(\Phi_k^{\pm}, 0, 0\right)^{\mathsf{T}}$, this relation can be written more compactly in the form

$$U_{k+1}^{\pm} = M_k U_k^{\pm} + \Phi_k^{\pm}. \tag{58}$$

The form of the solution of a linear difference equation of this type is known. It reads

$$U_{k+1}^{\pm} = M_k M_{k-1}...M_1 U_1^{\pm} + \sum_{l=1}^{k-1} M_k M_{k-1}...M_l \Phi_l^{\pm} + \Phi_k^{\pm}. \tag{59}$$

Based on this final result, it can be concluded that the Fuchsian differential equation (39) with five regular singular points has a second solution $F_2 = F_2(\xi)$ with coefficients resulting from (40), (43), (53), (56) and (59).

It may be noted that in this context, the application of the companion matrix approach proves to be particularly simple and therefore beneficial, but not absolutely necessary, since the solutions of linear recurrence relations of the form (48) and (56) can actually be written down in different ways [20, 28].

Nevertheless, the form of the coefficients used for the definition of $F_1^{\pm} = F_1^{\pm}(\xi)$ and $F_2^{\pm} = F_2^{\pm}(\xi)$ proves to be quite complicated regardless of the chosen approach anyway, which, however, will not cause any problems in the follwoing, since the exact form of said coefficients will only be of minor importance for the further course of the work.

What will turn out to be more important, however, is the fact that two exact solutions $F_1(\xi)$ and $F_2(\xi)$ of the Fuchsian differential equation (45), which is idential to (46), are now known. These solutions are power series given in terms of coefficients $w_k$ and $u_k$, which are solutions of recurrence relations of the form (50) and (58), whose solutions are (51) and (59). These solutions belong to a very large class of solutions, i.e. a class of not less than 1920 solutions, as can be concluded from the fact that the symmetry group acting on solutions of Fuchsian differential equations is isomorphic to the so-called Coxeter group $D_n$ of order $n!2^{n-1}$.

Having determined the exact form of two representatives of this large class, one can now make the ansatz of the form

$$F = \Theta_+ F^+ + \Theta_- F^- \tag{60}$$

for the reduced profile function $F$ of the geometry, where

$$F^\pm = c_1^\pm F_1^\pm + c_2^\pm F_2^\pm \tag{61}$$

and $\Theta_\pm(\xi) = \Theta(\pm\xi)$ is the Heaviside step function.

Since it is required that the solution is regular at $\xi_0 = -1$ for fixed $j$, it is known that the coefficient $c_2^-$ must be identically zero. Accordingly, using the fact that one can always set w.l.o.g. $\Theta_+ + \Theta_- = 1$, the solution can be expressed in the form

$$F = \Theta(c_1^+ F_1^+ + c_2^+ F_2^+) + (1 - \Theta)c_1^- F_1^-. \tag{62}$$

Note that the individual parts of this solution can be related w.l.o.g. by means of a monodromy transformation of the type

$$\begin{pmatrix} F_1^- \\ F_2^- \end{pmatrix} = \begin{pmatrix} a_{11} & a_{12} \\ a_{21} & a_{22} \end{pmatrix} \begin{pmatrix} F_1^+ \\ F_2^+ \end{pmatrix}, \tag{63}$$

which yields $F_1^- = a_{11}F_1^+ + a_{12}F_2^+$. This can be used to ensure that the conditions

$$c_1^+ F_1^+|_{\xi=0} + c_2^+ F_2^+|_{\xi=0} \stackrel{!}{=} c_1^- F_1^-|_{\xi=0}, \tag{64}$$

$$c_1^+ \frac{dF_1^+}{d\xi}\Big|_{\xi=0} + c_2^+ \frac{dF_2^+}{d\xi}\Big|_{\xi=0} = c_1^- \frac{dF_2^+}{d\xi}\Big|_{\xi=0} \tag{65}$$

both are fulfilled, which, however, requires the conditions $c_2^+ - a_{12}c_1^- \stackrel{!}{=} 0$ and $c_1^+ - a_{11}c_1^-$ to hold in the present context, which is not much of a problem due to the fact that the coefficients $c_1^\pm$ and $c_1^-$ can be freely chosen.

Having obtained the general form of the reduced profile function $F = F(\xi)$, the next step is now to solve the distributional relation $[\boxtimes f, \cdot] = [2\pi d_0 \delta(\xi-1), \cdot]$. Considering equation (43) and using $\Sigma_+^{-1} = (\Gamma_+ \bar{\Gamma}_+)^{-1}$ in combination with the identity $x\delta(x) \approx 0$, where $\approx$ means association in a distributional sense, one finds

$$\Gamma_+ \boxtimes f \approx 2\pi d_0(r_+ + ia)\delta_N \Rightarrow \Sigma_+ \boxtimes f \approx 2\pi d_0 \Sigma_+ \delta_N \approx 2\pi d_0(r_+^2 + a^2)\delta_N. \tag{66}$$

Thus, to obtain a solution of equation (44), it is sufficient to show that

$$[(\Delta_{\mathbb{S}_2} + V)f, \cdot] = [2\pi d_0(r_+^2 + a^2)\delta_N, \cdot] \tag{67}$$

is consistently met. Using for this purpose the fact that $\boxtimes f = 0$ can be fulfilled using a reduced profile funtion $F$ of the form (62), one finds

$$\lim_{\epsilon \to 0} \int_{\Omega \setminus B_\epsilon} f(\Delta_{\mathbb{S}_2} - V)\varphi \omega_q = \lim_{\epsilon \to 0} \int_0^{2\pi} d\phi \int_{-1}^{1-\epsilon} d\xi f \left\{ \frac{1}{2} \frac{d}{d\xi} \left( (1-\xi^2) \frac{d\varphi}{d\xi} \right) + V\varphi \right\} =$$

$$= \pi \lim_{\epsilon \to 0} (1-\xi^2) \left\{ \frac{d\varphi}{d\xi} f - \varphi \frac{df}{d\xi} \right\} \Big|_{-1}^{1-\epsilon}, \tag{68}$$

where $\Omega$ represents a two-section of the exterior event horizon, $B_\epsilon$ represents a ball shaped region around the 'north pole singularity' and $\omega_q$ represents the corresponding area element. A careful analysis shows here that $\lim_{\epsilon \to 0} \left[ (1-\xi^2) \frac{d\varphi}{d\xi} f \right] \Big|_{1-\epsilon} \to 0$. Consequently, using the Taylor expansion $\varphi = \varphi(\xi-1) = \varphi(0) + (\xi-1)\varphi'(0) + ...$, one finds after a careful treatment of all relevant terms

$$\lim_{\epsilon \to 0} \int_{\Omega \setminus B_\epsilon} f(\Delta_{\mathbb{S}_2} + V)\varphi \omega_q = 2\pi d_0 (r_+^2 + a^2) \varphi(0), \tag{69}$$

provided the fact that the definition $d_0 = \frac{-2C_1 c_2^+ w_0^+}{r_+^2 + a^2}$ is used in the present context. Accordingly, it can be concluded that relation (37) is met for the choice

$$f(v, \xi, r) = e^{\kappa v} \Sigma_+^{-1} F(\xi - 1) \delta(r - r_+), \tag{70}$$

on the exterior event horizon $\mathcal{H}_+$, where $F(\xi) = \Theta(\xi) F^+(\xi-1) + [1 - \Theta(\xi)] F^-(-(\xi+1))$ is a composite solution of (46) obtained from gluing $F^\pm(\xi)$ together. This solution is a rather complicated power series in the angular variable $\xi$, which, however, does not cause any problems, since only the form of the linear coefficients $c_1^\pm$ and $c_1^-$ entering relations (61) and (62) are needed anyway to solve (44) exactly.

Consequently, it becomes clear that the resulting expression for the profile function (62) solves differential equation (44) exactly and that the resulting Kerr-Schild deformation of the Kerr-Newman metric therefore represents an exact, distributionally well-defined solution of Einstein's field equations.

In contrast to the Kerr-Newman background spacetime, however, the solution obtained is no longer a stationary black hole spacetime. Moreover, the exterior event horizon of the black hole is no longer a Killing horizon, but rather a singular hyperface distorted by the field of the gravitational shock wave. As it turns out, said hypersurface is a weakly isolated horizon in the sense of Ashtekar et al., that is, a non-expanding null hypersurface endowed with a WIH structure $(\mathcal{H}^+, [l])$. This can be seen as follows: Using the Kerr-Schild deformed null tetrad $(l^a, \tilde{k}^a, m^a, \bar{m}^a)$ and the associated null co-tetrad $(-\tilde{k}_a, -l_a, \bar{m}_a, m_a)$ as well as the projectors $\tilde{\iota}_b^a = -l^a \tilde{k}_b + m^a \bar{m}_b + \bar{m}^a m_b$ and $\iota_b^a = -l^a k_b + m^a \bar{m}_b + \bar{m}^a m_b$, the result

$$[\mathcal{L}_l, \tilde{\mathscr{D}}_a] l^b = 0 \tag{71}$$

can be obtained, where $\mathcal{L}_l$ is the Lie derivative with respect to $l^a$ on $\mathcal{H}^+$. This follows from the fact that $\tilde{\mathscr{D}}_a l^b = \mathscr{D}_a l^b = \omega_a l^b$, $\mathcal{L}_l \omega_a = 0$ and $\mathcal{L}_l l^b = 0$, where $\tilde{\mathscr{D}}_a = \tilde{\iota}_a^b \tilde{\nabla}_b$ and $\mathscr{D}_a = \iota_a^b \nabla_b$ are the covariant derivatives defined with respect to the induced Kerr-Schild deformed geometry and the induced background geometry on $\mathcal{H}_+$. However, the validity of (71) is the key criterion for existence of a weakly isolated horizon. On the other hand, as it turns out, the distorted exterior event horizon is not an isolated horizon, since the condition

$$[\mathcal{L}_l, \tilde{\mathscr{D}}_a] q^b = 0 \tag{72}$$

is not fulfilled for any $q^b \in T(\mathcal{H}_+)$ in the given context. In the case of an undistorted event horizon $\mathcal{H}_+$ living in an undistorted Kerr-Newman black hole background $(M, g)$, however, this would very well apply. From this, however, it can be concluded that the solution obtained is that of a deformed black hole spacetime, which has local properties inherently different from those of the Kerr-Newman background geometry.

In any case, the resulting geometry not only has a different structure, but also a different physical interpretation than Kerr-Newman geometry. This can be illustrated by treating the problem of what happens to a freely falling observer when he/she passes the event horizon, i.e. the singular hyperface of the null particle. For the sake of simplicity, said observer shall be assumed to be a geodesic observer whose world line has the unit tangent $u^a = \dot{x}^a$. Based on this hypothesis, one may consider the geodetic deviation equation

$$\delta \ddot{x}^a = \tilde{R}^a{}_{bcd} u^b u^c \delta x^d, \tag{73}$$

which, despite of the fact that the profile 'function' $f$ is proportional to Dirac's delta distribution, turns out to be well-defined in this context. This is not least because the Kerr-Schild deformed Riemann curvature tensor $\tilde{R}^a{}_{bcd}$ turns out to be linear in the profile function $f$. To see this, several steps have to be taken. In the first step, one may again apply methods of Colombeau's theory of generalized functions and consider a regularization $\delta_\varepsilon$ of the delta distribution and decompose the Kerr-Schild deformed Riemann curvature tensor in the form

$$\tilde{R}^a{}_{bcd} = R^a{}_{bcd} + \rho^a{}_{bcd}, \tag{74}$$

where $R^a{}_{bcd}$ is the curvature tensor of the Kerr-Newman background and $\rho^a{}_{bcd} = 2\nabla_{[b} C^a{}_{c]d} + 2 C^e{}_{b[d} C^a{}_{c]e}$ with $C^a{}_{bc} = \frac{1}{2} \left( \nabla_b (f l^a l_c) + \nabla_c (f l^a l_b) - \nabla^a (f l_b l_c) \right)$ is its Kerr-Schild deformed counterpart[2]. In addition, one may take the fact into account that $\nabla_a l^c = \omega_a l^c + l_a \omega^c$ holds locally on the horizon of the black hole and therfore $\nabla_a l^c \nabla_b l_c = \omega^2 l_a l_b$ and $\nabla_a l^c \nabla_c f = (\omega \mathcal{D}) f l_a$ as well. By repeated usage of these relations, one then finds that

$$C^e{}_{b[d} C^a{}_{c]e} = 0 \Rightarrow \rho^a{}_{bcd} = 2 \nabla_{[b} C^a{}_{c]d} \tag{75}$$

---

[2] Note that condition (9) was tacitly used at this point to obtain this particularly simple form of the difference tensor $C^a{}_{bc}$.

holds on $\mathcal{H}_+$ and therefore all over $(\tilde{M}, \tilde{g})$. Note that, by using conditions (13) and (20), a similar result has been obtained for the Ricci and Einstein tensors (regardless of the position of indices) in [21].

Based on these insights, one may use next the relation $\tilde{R}^a_{bcd} = \tilde{C}^a_{bcd} + \delta^a_{[c}\tilde{R}_{d]b} - \frac{1}{3}\tilde{R}\delta^a_{[c}\tilde{g}_{d]b}$ to re-write equation (73) in the following form

$$\delta\ddot{x}^a = \tilde{C}^a_{bcd}u^b u^c \delta x^d + \frac{1}{3}\tilde{R}_{db}u^d u^b \delta x^a - \frac{1}{2}\tilde{S}^a_c \delta x^c, \qquad (76)$$

where $\tilde{S}^a_c = \tilde{h}^{ad}\tilde{h}^b_c\tilde{R}_{db} - \frac{1}{3}\tilde{h}^a_c\tilde{h}^{db}\tilde{R}_{db}$ with $\tilde{h}_{ab} = \tilde{g}_{ab} + u_a u_b$. As first pointed by Szekeres [34], this relation can be used to show (via using the Petrov classification) that the geodesic deviation equation decomposes in the linear sum of three distinct components - a transverse wave, a longitudinal wave, and a Coulomb component, each having a characteristic effect on the geodesic congruence of timelike curves defined by the free-falling observer.

This insight can be used to make the following observation: By performing a Kerr-Schild deformation of the metric of the form (2), the geometry of spacetime changes in such a way that it is no longer Petrov type $D$. More specifically, the ambient spacetime $(\tilde{M}, \tilde{g})$ turns out to be a Petrov type $II$ spacetime. Consequently, however, considering a normalized geodesic frame of the form $(u^a, s^a, E^a_2, E^a_3)$ and its associated co-frame $(-u_a, s_a, e^2_a, e^3_a)$, where $u^a = \frac{1}{\sqrt{2}}\left(l^a + \tilde{k}^a\right)$, $s^a = \frac{1}{\sqrt{2}}\left(l^a - \tilde{k}^a\right)$, $E^a_2 = \frac{1}{\sqrt{2}}(m^a + \bar{m}^a)$ and $E^a_3 = \frac{i}{\sqrt{2}}(m^a - \bar{m}^a)$, equation (73) yields two relations of the form

$$\delta\ddot{x}^a = \tilde{\Psi}_2[s^a s_c - \frac{1}{2}(E^a_2 e^2_c + E^a_3 e^3_c)]\delta x^c \qquad (77)$$

and

$$\delta\ddot{x}^a = \frac{1}{2}\tilde{\Psi}_4(E^a_2 e^2_c - E^a_3 e^3_c)\delta x^c. \qquad (78)$$

At first sight, both of these equations look extremely problematic and mathematically completely ill-defined due to the fact that $\tilde{\Psi}_2 = \tilde{C}_{abcd}l^a m^b \bar{m}^c \tilde{k}^d$ and $\tilde{\Psi}_4 = \tilde{C}_{abcd}\tilde{k}^a \bar{m}^b \tilde{k}^c \bar{m}^d$. However, using Colombeau's theory of generalized functions once again, i.e., considering once more a delta sequence $\delta_\varepsilon$ which coincides with the delta distribution in the Colombeau limit $\varepsilon \to 0$, one finds surprisingly that both relations (77) and (78) are indeed well-defined as long as it is required that $l_c \delta x^c \stackrel{!}{=} 0$. To see this, one may check that the relations $C^a_{bc}l_a = 0$ and $\nabla_a l^c = \omega_a l^c + l_a \omega^c$ imply that $\rho^a_{bcd}l_a = \omega_a C^a_{d[b}l_{c]}$ and therefore $\tilde{\Psi}_2 \approx \Psi_2$ holds on the horizon of the black hole. Finally, using then the relations $\tilde{C}_{abcd}l^a m^b \bar{m}^c l^d = \tilde{C}_{abcd}l^a m^{(b}\bar{m}^{c)}l^d = \tilde{C}_{abcd}l^a \left(\frac{1}{2}g^{bc} + l^{(b}k^{c)}\right)l^d = 0$ and $\tilde{C}_{abcd}l^{(a}k^{c)}\bar{m}^b \bar{m}^d = \tilde{C}_{abcd}\left(-\frac{1}{2}g^{bc} + m^{(b}\bar{m}^{c)}\right)\bar{m}^b \bar{m}^d = 0$, one finds after taking the limit limit $\varepsilon \to 0$ that both (77) and (78) remain linear in $f$ and therefore are indeed well-defined.

But unfortunately this does not save our observer: Due to the fact that both the Coulomb component (77) and the transversal wave component (78) of the geodesic deviation equation (76) cointain terms that are proportional to $\delta(0)$, the distortion resulting from the tidal force acting on an infalling geodesic observer turns out to be infinite. This allows one to draw the following conclusion: An observer who survives the free fall through the event horizon of a Kerr-Newman black hole does by far not survive the fall through (or rather the collission with) a particle shock wave on the horizon of a deformed Kerr-Newman black hole. In this respect, the shock wave acts like a firewall that tears apart all objects that threaten to fall into the black hole with infinite tidal forces. In this sense, the properties of the geometry obtained in this work and the associated Kerr-Newman background geometry are locally quite different.

## 3 Geometric Limits and Physical Interpretation

As already mentioned in section one of this work, the generalized Dray-'t Hooft relation reduces to Sfetsos' relation in the limit $a \to 0$ and to the Dray-'t Hooft relation in the limit $a, e \to 0$. In the more general case of the limit $a \to 0$, one is left to solve the much simpler problem

$$[\boxtimes f, \cdot] = [2\pi b_0 \delta(\xi - 1), \cdot], \tag{79}$$

where $\boxtimes = \frac{d}{d\xi}((1 - \xi^2)\frac{d}{d\xi}) - c$. Making an ansatz of the form

$$f(v, \xi, r) = e^{\kappa v} F(\xi - 1)\delta(r - r_+), \tag{80}$$

where $\alpha$ and $\beta$ are constants, this equation can be solved by considering first the homogeneous equation

$$\boxtimes F = \frac{d}{d\xi}((1 - \xi^2)\frac{dF}{d\xi}) - cF = 0, \tag{81}$$

which obviously matches Legendre's differential equation in case that the constant $c$ can be written in the form $c = -l(l + 1)$. In this comparatively simple case, the said equation admits two independent solutions known as Legendre functions of first and second kind, which result as a special case of Gauß's differential equation [33]. These functions may therefore be expressed in the form $F_1(\xi) = F(l + 1, -l, 1; \frac{1-\xi}{2}) =: P_l(\xi)$ and $F_2(\xi) = \frac{1}{2}P_l(\xi)\ln|\frac{1+\xi}{1-\xi}| - \sum_{k=1}^{L}\frac{2l-4k+3}{(2l-1)(l-k+1)}P_{l-2k+1}(\xi)$, where $L = \frac{1}{2}l$ applies in the case that $l$ is even and $L = \frac{1}{2}(l + 1)$ applies in the case that $l$ is odd. Of course, by making a transformation of the type $\xi \to -\xi$, different pairs of solutions $F_1^\pm = F_1^\pm(\xi)$ and $F_2^\pm = F_2^\pm(\xi)$ are obtained. These individual pairs of solutions can be glued together completely the same way as previously shown in the more general axisymmetric case, which yields the reduced profile function

$$F = \Theta_+ F^+ + \Theta_- F^-, \tag{82}$$

where

$$F^\pm = c_1^\pm F_1^\pm + c_2^\pm F_2^\pm. \tag{83}$$

Thus, using the same definitions as in the axisymmetric case, this solution can be re-written in the form

$$F = \Theta(c_1^+ F_1^+ + c_2^+ F_2^+) + (1-\Theta)c_1^- F_1^-. \tag{84}$$

This yields the conditions

$$c_1^+ F_1^+|_{\xi=0} + c_2^+ F_2^+|_{\xi=0} \stackrel{!}{=} c_1^- F_1^-|_{\xi=0} \tag{85}$$

$$c_1^+ \frac{dF_1^+}{d\xi}|_{\xi=0} + c_2^+ \frac{dF_2^+}{d\xi}|_{\xi=0} = c_1^- \frac{dF_2^+}{d\xi}|_{\xi=0} \tag{86}$$

which, of course, can also be met in the given spherically symmetric case. The reduced profile function $F$ therefore fulfills the Sfetsos' relation given above in the case that $c = -l(l+1)$.

However, since it may not always be possible to assume the validity of that relation, it may be necessary to proceed differently. As was first demonstrated by Dray and 't Hooft, a particular way to do so is to solve the inhomogeneous equation directly, which can be achieved by expanding the reduced profile function on the left hand side and the delta function on the right hand side simultaneously in Legendre polynomials. Since it is known that $\delta(x) = \sum\limits_{l=0}^{\infty}(l+\frac{1}{2})P_l(x)$, one obtains the solution

$$F(\xi) = -b_0 \sum_{l=0}^{\infty} \frac{l+\frac{1}{2}}{l(l+1)+c} P_l(\xi) \tag{87}$$

by solving the corresponding eigenvalue problem. An integral expression for this solution can be found by considering the generating function of the Legendre polynomials

$$\sum_{l=0}^{\infty} \frac{l+\frac{1}{2}}{l(l+1)+c} P_l(\xi) e^{-sl} = \frac{1}{\sqrt{1-2\xi e^{-s}+e^{-2s}}}$$

in addition to the fact that

$$\frac{l+\frac{1}{2}}{l(l+1)+\alpha^2+\frac{1}{4}} = \int_0^{\infty} e^{-s(l+\frac{1}{2})} \cos(\alpha s) ds.$$

This yields

$$F(\xi) = -\frac{b_0}{\sqrt{2}} \int_0^{\infty} \frac{\cos(\sqrt{c-\frac{1}{4}}s)}{\sqrt{\cosh s - \xi}} ds. \tag{88}$$

While handling the whole subject this way obviously works well enough, there is actually another way to proceed in this regard. One may simply solve the corresponding differential equation directly without relying on the existence of polynomial solutions. This can be seen as follows: Starting with an ansatz of the form $F_1 = \sum_{k=0}^{\infty} w_k(\xi-1)^k$ and $F_2 \equiv F_1\Delta + G$, according to which $G = \sum_{k=0}^{\infty} u_k(\xi-1)^k$ applies, one obtains two different solutions of Sfetsos' differential relation under the condition that the corresponding coefficients fulfill $w_{k+1} = \prod_{j=0}^{k} m_j \cdot w_0$ with $m_j = m(j) = -\frac{j(j+1)+c}{2(j+1)^2}$ and $u_{k+1} = \prod_{j=0}^{k} m_j \cdot (u_0 + \sum_{j=0}^{k} \frac{D_1 w_0}{k-j+1})$ and the logarithmic part of the Green function is given by $\Delta = \frac{D_1}{2}\ln|\frac{1+\xi}{1-\xi}| + D_2$, where $D_1, D_2$ are arbitrary constants. By performing then once again a transformation of the type $\xi \to -\xi$, different pairs of solutions $F_1^{\pm} = F_1^{\pm}(\xi)$ and $F_2^{\pm} = F_2^{\pm}(\xi)$ are obtained. These individual pairs of solutions can be glued together completely the same way as previously shown. This yields once again a reduced profile function of the form

$$F = \Theta(c_1^+ F_1^+ + c_2^+ F_2^+) + (1-\Theta)c_1^- F_1^-, \qquad (89)$$

which, apart from a slightly different form of the integration constants, is exactly what is obtained by considering the limit $a \to 0$ of the previously obtained solution of the generalized Dray-'t Hooft equation. Conversely, using the fact that the coefficients $m_{k,1} = -\frac{r_+^2}{2(r_+^2+a^2)}\frac{k(k+1)}{(k+1)^2} - \frac{a^2}{2(r_+^2+a^2)}\frac{(5k-3)k}{(k+1)^2} - \frac{\kappa r_+}{(k+1)^2}$, $m_{k,2} = -\frac{2a^2}{r_+^2+a^2}\frac{(k-2)(k-1)}{(k+1)^2}$ and $m_{k,3} = -\frac{a^2}{r_+^2+a^2}\frac{(k-2)(k-3)}{(k+1)^2}$ deduced in the previous section reduce to the form $m_{k,1} = -\frac{k(k+1)+2\kappa r_+}{2(k+1)^2}$, $m_{k,2} = 0$ and $m_{k,3} = 0$ in the limits $a \to 0$ and/or $a, e \to 0$, one finds that the matrix recurrence relations (51) and (59) acquire an extremely simple form and become solvable af trivial manner. Consequently, it can be concluded that the solution for the reduced profile function (62) of generalized Dray-'t Hooft equation reduces exactly to Sfetsos' solution in the limit $a \to 0$ and to the original solution of Dray and 't Hooft in the limit $a, e \to 0$, both of which can be characterized by a solution for the reduced profile function of the form (89).

Accordingly, since this expression and the one previously obtained both solve the corresponding differential equations (35) and (36), it becomes clear that they must be identical in the limit $e \to 0$. Therefore, it can be concluded that they are different expressions of one and the same reduced profile function of the geometry.

As can be seen, the solution obtained differs in several ways from the solutions of Dray and 't Hooft and Sfetsos. This is not particularly surprising in that the rotation one-form $\omega_b$ does not occur in the spherical equations, with the consequence being that the solutions of the generalized Dray-'t Hooft equation are no longer hypergeometric functions as in the case of the Dray-'t Hooft equation or the Sfetsos equation, but generalizations of such. The reason for this is the following: Considering the results of Jaramillo's work, it becomes clear that the rotation one-form $\omega_b$ has similar physical signficance for the form of differential equation (26) as the gauge field $A_b$ for the Schrödinger equation (33) of a charged quantum particle. For if the rotation is allowed to approach zero in a suitable limit, a differential equation is obtained which can be interpreted as a Schrödinger equation for an uncharged quantum particle, no longer as that for a charged particle. From this, however, it can be concluded that the angular momentum of the black hole in the present case plays a completely analogous role to that of the charge in the quantum particle case, which is why it immediately becomes clear why the solutions of the respective differential equations in the rotating and non-rotating case have such a different form. In contrast, the vanishing of the black hole charge does not have such a significant influence on the structure of the generalized Dray-'t Hooft equation and therefore also not on the form of the corresponding solutions.

Since the geometry obtained, like the Kerr-Newman solution, depends on three physical parameters, it is immediately clear that it defines a whole family of gravitational shock wave spacetimes. The representatives of this family are given by black hole spacetimes deformed by a gravitational shock wave generated by a null particle located at the exterior event horizon, which are rare examples of superimposed Kerr-Schild deformed two-body solutions Einstein's field equations. The exterior event horizon of these spacetimes is no longer a Killing horizon but an extremely weakly isolated horizon in the sense of Ashtekar et al. Moreover, the geometry is no longer stationary and axisymmetric, so it can legitimately be interpreted as the field of a dynamical black hole, whose dynamical components are given solely by the field of a point particle (or a distribution of point particles) that envelops the exterior event horizon, thus permanently altering its intrinsic geometric structure.

Due to the logarithmic singularity that occurs in the definition of (62), the mentioned class of gravitational shock wave spacetimes has a very peculiar and unique causal structure that can be used as a template for studying the behavior of spacetime metrics with low regularity. This is not least because said class of solutions has geometric properties inherently different from those of pp-wave spacetimes and other distributional spacetimes, and it turns out to be a specific representative of an infinite set of solutions with exactly the same physical interpretation but with different mathematical structures (some of which may be simple). This can be seen as follows: By performing a rescaling of the form $l^a \to \lambda l^a$, $k^a \to \lambda^{-1} k^a$, where $\lambda = \lambda(\xi)$ is some smooth non-vanishing function, one obtains a generalized Kerr-Schild metric of the form

$$\tilde{\tilde{g}}_{ab} = g_{ab} + \lambda^2 f l_a l_b, \tag{90}$$

which, if $D\lambda \stackrel{!}{=} 0$ is met in the present context, is again a null fluid solution of the deformed field equations $\tilde{\rho}^a_{\ b} = 8\pi \tilde{\tau}^a_{\ b}$, where $\tilde{\rho}^a_{\ b} = -\tilde{\boxtimes} \tilde{f} \tilde{l}^a \tilde{l}_b = -\lambda^2 \tilde{\boxtimes} \tilde{f} l^a l_b$ with

$$\begin{aligned}\tilde{\boxtimes} :=& \frac{1}{2}\mathcal{D}^2 + (2(\tilde{\alpha} + \tilde{\tilde{\beta}}) - \tilde{\tilde{\tau}})\delta + (2(\tilde{\alpha} + \tilde{\beta}) - \tilde{\tau})\delta' + \frac{1}{2}\tilde{D}'(\tilde{\rho} + \tilde{\rho}) + \\ &+ \delta(\tilde{\alpha} + \tilde{\tilde{\beta}} - \tilde{\tilde{\tau}}) + \delta'(\tilde{\tilde{\alpha}} + \tilde{\beta} - \tilde{\tau}) + (\tilde{\tilde{\tau}} + \tilde{\alpha} - \tilde{\tilde{\beta}})(\tilde{\tau} - \tilde{\tilde{\alpha}} - \tilde{\beta}) + \\ &+ (\tilde{\tau} + \tilde{\tilde{\alpha}} - \tilde{\beta})(\tilde{\tilde{\tau}} - \tilde{\alpha} - \tilde{\tilde{\beta}}) + 4|\tilde{\alpha} + \tilde{\tilde{\beta}}| - \tilde{\tilde{\tau}}(\tilde{\alpha} + \tilde{\beta}) + \\ &- \tilde{\tau}(\tilde{\alpha} + \tilde{\tilde{\beta}}) + \tilde{\Psi}_2 + \tilde{\tilde{\Psi}}_2. \end{aligned} \tag{91}$$

Thus, given the form (70) of the profile function entering the Kerr-Schild deformed Kerr-Newman metric (2), one can easily obtain a new Kerr-Schild deformed metric with similar structure and physical interpretation. By continuous rescaling, one can then easily obtain an infinite class of solutions of Einstein's field equations, all of which have similar (but not identical) geometric properties and physical interpretations. Since the field equations are linear, all these solutions can be superimposed in order to obtain a general relativitsitc manybody solution for a distribution of null particles located at the black hole event horizon.

Finally, it shall be emphasized that the solution found could have a potentially interesting application in that the corresponding shock wave geometry could describe, in principle, the gravitational backreaction caused by Hawking quanta that are emitted to null infinity along the event horizon of a Kerr-Newman black hole by the effect of Hawking evaporation. After all, there are works in the literature [25, 26] which intend to show that the shock wave geometry constructed by Dray and 't Hooft characterizes such a backreaction in the much simpler case of a spherically symmetric Schwarzschild black hole. But since the solution obtained in this work is an extension of the Dray-'t Hooft geometry, it should of course have a similar physical interpretation at the end of the day. Yet, the situation seems to be such that until today there is no generally accepted approach that relates the theory of shock wave spacetimes to the theory of scalar fields on stationary black hole backgrounds. In this respect, it is worth being cautious with such interpretation attempts.

## Discussion

In the present work, the gravitational field of a shockwave generated by a massless particle was calculated at the event horizon of a stationary Kerr-Newman black hole. The resulting solution of Einstein's equation represents the local geometric field of a relativistic two-body system (local in the sense of [22]), which may in principle be extended to a many-body system by superimposing different expressions for the profile function of the geometry.

More precisely, using the Kerr-Schild metric deduced in this work, another local energy-momentum distribution of a null particle located on the event horizon of a black hole can be determined by introducing a suitable null rescaling, so that another Kerr-Schild geometry is obtained that does not match the original one, but yields the same type of energy-momentum tensor of a point-like null particle located at the event horizon of spacetime. This provides an infinite set of solutions of Einstein's equations, which, despite the fact that they have a different geometric structure, are all equal from a physical point of view. Accordingly, since the Einstein tensor is linear in the profile function, a finite series of these solutions for the reduced profile function can in principle be superimposed to a many-body solution, although the solution in question hardly turns out to be physically relevant in that all null particles move exactly in the same direction with exactly the same velocity, which appears to be too much of an idealization. Nevertheless, such a solution can in principle be written down.

A necessary prerequisite for doing so is, of course, the calculation of the exact form of the profile function of the geometry on the level of infinite power series. Regarding this exact form, the results presented in this work differ greatly from those of a prior work on the subject presented by BenTov and Swearngrin [11], which, like this work, is dedicated to the the construction of a Kerr-Schild shockwave in a Kerr-Newman background, but which failed on the task of providing an exact expression for the profile function of the geometry (despite its claim to have found an exact solution to Einstein's equation). However, as must be acknowledged, the geometric setting considered in this work differs insofar considerably from that considered in [11] as a different choice for the coordinates of the background metric was made. Consequently, since both works set up deformations in different charts of Kerr-Newman spacetime, the results of both approaches can hardly be meaningfully compared with each other.

In any case, the geometric model constructed in the present work provides precise information on the profile function of the geometry and thus allows for the speficiation of an entire family of superimposed generalized Kerr-Schild spacetimes in four dimensions, which proves to be consistent with models already existing in the literature. In particular, it contains the solutions of Dray and 't Hooft, Sfetsos and, of course, Aichelburg and Sexl as a special case. In exactly the same way as in all these spacetime models, the profile function results from suitable gluing of individual solutions of the field equations, whereas in the given Kerr-Newman case (in contrast to the mentioned spherically symmetric predecessor models) so far only gluing of power series solutions, not polynomial solutions, turns out to be possible. It would certainly be interesting to address this shortcoming in a future extension of the model. In addition, it would be desirable to apply the presented method for calculating the metrics of shock wave spacetimes to other stationary background spacetimes such as the Kerr-Newman-Taub-NUT spacetime, or even to apply the associated deformation method to simple non-stationary models such as the Kerr-Newman-Vaidya spacetime. After all, there is potential that in this way new candidates for distributional spacetimes can be found, whose geometric structures can all be derived - in agreement with the results of [21] - from the Hamiltonian principle.

**Acknowledgements:**

I want to thank Herbert Balasin, Walter Simon and Andi Kastner for interesting discussions on the subject.